\documentstyle[12pt]{article}
\def\shiftdown#1{#1\llap{\lower.04ex\hbox{#1}}}

\input epsf.sty

\begin{document}

\author{S. Kullander$^a$,  H. Cal\'en$^a$, K. Fransson$^a$, A.
Kup\'s\'c$^{a,b}$, \\B.~Martemyanov$^d$,
B.~Morozov$^{a,c}$,R.J.M.Y.~Ruber$^a$, V.~Sopov$^d$,\\ 
\vspace{3mm}
J.~Stepaniak$^b$,
V.~Tchernychev$^d$\\
$^a$Department of Radiation Sciences, University of Uppsala,\\ Box
535,
S-75121 Uppsala, Sweden;\\
$^b$Institute for Nuclear Studies, PL-00681 Warsaw, Poland;\\
$^c$Joint Institute for Nuclear Research, Dubna, \\101000 Moscow, Russia;\\
$^d$Institute for Theoretical and Experimental Physics, \\117259 Moscow,
Russia.}

\title{Rare decays of $\eta$-meson: the background of
$\rho$- and $\omega$- mesons }

\maketitle

\begin{abstract}
The problem of $\rho$- and $\omega$- mesons contributions to the
background for $\eta \rightarrow \pi^0 e^+e^-$ decay and to the
asymmetries in the decays $\eta \rightarrow \pi^+\pi^-\pi^0$
and  $\eta \rightarrow \pi^+\pi^-\gamma$ is considered.
\end{abstract}
\section{Introduction}
 The reaction $pp \rightarrow pp\eta$ at the proton energy
$E_p = 2.3 GeV$ ($\sqrt{s} = 2.47 GeV$) has the cross section about
$5 \mu b$  ~\cite{WASA1,WASA2}.  Then for the luminosity of WASA at
CELSIUS about $10^{32} cm^2/sec$ ~\cite{WASA1,WASA2} $500$ $ \eta$- mesons will
 be produced
every second. This means that the reaction $\eta \rightarrow \pi^0 e^+e^-$
with the expected branching ratio $$Br(\eta \rightarrow \pi^0 e^+e^-) \sim 10^
{-9} ~~\cite{Ng1,Ng2}$$ \\
and the asymmetries in the reactions
$\eta \rightarrow \pi^+\pi^-\pi^0$ and $\eta \rightarrow \pi^+\pi^-\gamma$
with the expected level
$$A(\eta \rightarrow \pi^+\pi^-\pi^0)\sim 10^{-5}$$
$$A(\eta \rightarrow \pi^+\pi^-\gamma)\sim 10^{-5}~~ \cite{?}$$
will become accessible for investigations after a year of running
(approximately $10^{10} $ $\eta$'s).

Taking into account the expected level of the effects
there should be the strong restrictions on the background
processes. In this note we are considering the background decays
of $\rho$- and $\omega$- mesons that can be produced in the
$pp$ -collisions at $E_p = 2.3 GeV$ under the threshold. The relevant
parameters of $\rho$- and $\omega$- mesons  decays are the following 
 ~\cite{PDG}

$$Br(\rho \rightarrow \pi^0 e^+e^-) \approx 4*10^{-6}$$
$$Br(\omega \rightarrow \pi^0 e^+e^-) \approx 6*10^{-4}$$
$$Br(\omega \rightarrow \pi^+ \pi^-\pi^0) \approx 0.9$$
$$Br(\rho \rightarrow \pi^+ \pi^-\gamma) \approx 10^{-2}$$

In the next section we will evaluate the cross sections of
$\rho$- and $\omega$- mesons production and the corresponding
background in the considered decay channels of $\eta$- meson.

\section{Under the threshold production of$\rho$- and $\omega$- mesons  }
We will evaluate the cross sections of
$\rho$- and $\omega$- mesons production  in the region of invariant masses
($\sqrt{q^2}$) near the mass of $\eta$- meson ($m_{\eta} \pm \frac{\Delta m}{2},
\Delta m = 10 ~MeV$).
We will describe the production of $\rho$- and $\omega$- mesons by the
diagrams Fig.1
where the vertecies of $pp$- interaction and vector meson emission
are considered as constants, i.e. independent on the energy of the
incident proton ($E_p$) and the invariant mass of emitted meson ($\sqrt{q^2}$).

Then the production cross section is defined by the accessible phase space
factors, the resonance form of $\rho$- and $\omega$- mesons propagators 
and the $q^2$- dependence of $\rho$- and $\omega$- mesons decay widths.
Let us therefore write the differential production cross section in
the following form
\begin{equation}
\label{dcs}
\frac{d\sigma_V}{dq^2}=\sigma_V(E_p,q^2)\frac{1}{((q^2-m_V^2)^2+m_V^2\Gamma_
V^2(q^2))}
\frac{m_V\Gamma_V(q^2)}{\pi}~,
\end{equation}
where $\sigma_V(E_p,q^2)$ is the production cross section  of
a particle with invariant mass $\sqrt{q^2}$ that depends on $E_p$ and  $q^2$
only due to accessible phase space (approximation),\\
$\Gamma_V(q^2)$ is the $q^2$- dependent width of the resonance.

In oder to get the desireable cross sections of  $\rho$- and $\omega$- mesons
production the differential cross section (\ref{dcs}) should be integrated
over $q^2$ in the region $(m_\eta-5 MeV)^2 < q^2 < (m_\eta+5 MeV)^2$.
The unknown constant in  $\sigma_V(E_p,q^2)$ can be found using experimental
data on $\rho$- and $\omega$- mesons production at energies $E_p$ above
the threshold (see later).

Let us now describe the $q^2$- dependence of $\rho$- and $\omega$- mesons
 decay widths  $\Gamma_V(q^2)$. For $\rho$- meson we will take into account
the two particle phase space ($\rho \rightarrow \pi\pi$ is the main decay
channel of $\rho$- meson) and the $p$-wave character of
$\rho \rightarrow \pi\pi$ decay
\begin{equation}
\label{gammarho}
\Gamma_\rho (q^2) \sim \sqrt{\frac{q^2-4m_\pi^2}{q^2}}(q^2-4m_\pi^2)~.
\end{equation}

For $\omega$- meson the main decay channel is $\omega\rightarrow \pi\pi\pi$. The
decay width can be calculated numerically using the following form
 of matrix element (uniquely determined)
\begin{equation}
\label{gammaomega}
M(\omega\rightarrow \pi\pi\pi)\sim \epsilon_{\mu\nu\alpha\beta}
\epsilon_\mu p_{1\nu}p_{2\alpha}p_{3\beta}\frac{1}{q^3}~,
\end{equation}
where $\epsilon_\mu$ is the polarization vector of $\omega$- meson
and $p_i$ are the momenta of pions.

The $\rho$- meson production cross section is known at $\sqrt{s}= 3.5 GeV$
~\cite{rhoprod} and is equal to $\approx 80 \mu b$. Recalculating it at
the $\sqrt{s}= 2.47 GeV$ and   $(m_\eta-5 MeV)^2 < q^2 < (m_\eta+5 MeV)^2$
we get
$$\sigma_\rho = \frac{80}{2*10^5} = 4*10^{-4} \mu b ~.$$
The $\omega$- meson production cross section is known at $\sqrt{s}= 3.08 GeV$
~\cite{omegaprod} and is equal to $\approx 80 \mu b$. Recalculating it at
the $\sqrt{s}= 2.47 GeV$ and   $(m_\eta-5 MeV)^2 < q^2 < (m_\eta+5 MeV)^2$
we get
$$\sigma_\omega = \frac{80}{5*10^6} = 1.6*10^{-5} \mu b~.$$

Using the calculated cross sections and the branching ratios of
$\rho$- and $\omega$- decays mentioned in the Introduction we get
$\rho$- and $\omega$- backgrounds for $\eta \rightarrow \pi^0e^+e^-$
decay
$$\frac{\sigma_\rho Br(\rho \rightarrow \pi^0e^+e^-)}
{\sigma_\eta Br(\eta \rightarrow \pi^0e^+e^-)} = 0.32$$
$$\frac{\sigma_\omega Br(\omega \rightarrow \pi^0e^+e^-)}
{\sigma_\eta Br(\eta \rightarrow \pi^0e^+e^)} = 2$$
and the following background in the asymmetries \cite{Uto} of
$\eta \rightarrow \pi^+\pi^-\pi^0$
and  $\eta \rightarrow \pi^+\pi^-\gamma$- decays
$$A_\omega(\eta \rightarrow \pi^+\pi^-\pi^0)=
\sqrt{\frac{2\pi\Gamma_\eta}{\Delta m}
\frac{\sigma_\omega Br(\omega \rightarrow \pi^+\pi^-\pi^0)}
{\sigma_\eta Br(\eta \rightarrow \pi^+\pi^-\pi^0)}}\approx 10^{-4}$$
$$A_\rho(\eta \rightarrow \pi^+\pi^-\gamma)=
\sqrt{\frac{2\pi\Gamma_\eta}{\Delta m}
\frac{\sigma_\rho Br(\rho \rightarrow \pi^+\pi^-\gamma)}
{\sigma_\eta Br(\eta \rightarrow \pi^+\pi^-\gamma)}}\approx 10^{-4}~.$$\\
Here $\Delta m$ is equal to $10 MeV$. The asymmetries in both cases
are due to the charged pions in the background amplitudes have the opposite
symmetry under the intergange of momenta comparing to the  chaged pions
from the $\eta$- decays amplitudes and the amplitudes of the process and of
the background interfere.

\section{Conclusion}
 
So, we have considered the 
problem of $\rho$- and $\omega$- mesons contributions to the
background for $\eta \rightarrow \pi^0 e^+e^-$ decay and to the
asymmetries in the decays $\eta \rightarrow \pi^+\pi^-\pi^0$
and  $\eta \rightarrow \pi^+\pi^-\gamma$. We have found that the
background  due to  $\rho$- and $\omega$- mesons is comparable
with the expected signal from $\eta$- meson decay. For the case of
$\eta \rightarrow \pi^0 e^+e^-$ decay the background can be lowered
by lowering $\Delta m$  (better energy resolution) or by selecting the
events with the invariant mass of $e^+e^-$- pair in the region where
the decays of vector mesons do not substantially contribute.
In the case of asymmetries the first way (lowering of $\Delta m$  )
will not change the situation because of the effect (interference with
the background) does not depend on $\Delta m$. So, in this case
there should be looked for the regions on the Dalitz plot where
the ratio of the signal to background is higher. Now this work is in progress.

\newpage
{\bf Figure caption} {\bf Fig.1} {Diagrams of $\rho$ and $\omega$ production}

\end{document}